\documentclass{acm_proc_article-sp}

\usepackage{xspace}
\usepackage{url}
\usepackage{tabularx}
\usepackage{multirow}
\usepackage{color}
\definecolor{darkred}{rgb}{0.44,0,0}
\definecolor{darkgreen}{rgb}{0,0.44,0}
\definecolor{darkblue}{rgb}{0,0,0.44}
\definecolor{grey}{rgb}{0.5,0.5,0.5}

\begin{document}

\newcommand{\gemm}{{\sc gemm}\xspace}

\title{Performance and Energy Optimization of Matrix \\ Multiplication
       on Asymmetric big.LITTLE Processors}
%
%
%
%
%

\numberofauthors{6} 
%
\author{
%
%
\alignauthor
Sandra Catal\'an\\
       \affaddr{Depto. Ingenier\'ia y Ciencia de Computadores}\\
       \affaddr{Universitat Jaume~I, Castell\'on, Spain}\\
       \email{catalans@uji.es}
\alignauthor
Francisco D. Igual\\
       \affaddr{Depto. Arquitectura de Computadores y Autom\'atica}\\
       \affaddr{Universidad Complutense de Madrid, Spain}\\
       \email{figual@ucm.es}
\alignauthor
Rafael Mayo\\
       \affaddr{Depto. Ingenier\'ia y Ciencia de Computadores}\\
       \affaddr{Universitat Jaume~I, Castell\'on, Spain}\\
       \email{mayo@uji.es}
\and
\alignauthor
Luis Pi\~nuel\\
       \affaddr{Depto. Arquitectura de Computadores y Autom\'atica}\\
       \affaddr{Universidad Complutense de Madrid, Spain}\\
       \email{lpinuel@ucm.es}
\alignauthor
Enrique S. Quintana-Ort\'{\i}\\
       \affaddr{Depto. Ingenier\'ia y Ciencia de Computadores}\\
       \affaddr{Universitat Jaume~I, Castell\'on, Spain}\\
       \email{quintana@uji.es}
\alignauthor
Rafael Rodr\'{\i}guez-S\'anchez\\
       \affaddr{Depto. Ingenier\'ia y Ciencia de Computadores}\\
       \affaddr{Universitat Jaume~I, Castell\'on, Spain}\\
       \email{rarodrig@uji.es}
}

\maketitle
\begin{abstract}
Asymmetric processors have emerged as an appealing technology for severely energy-constrained
environments, especially in the mobile market where heterogeneity in applications is 
mainstream. In addition, given the growing interest on ultra low-power architectures for 
high performance computing, this type of platforms are also being investigated in the road towards the implementation of energy-efficient high-performance scientific
applications. In this paper, we propose a first step towards a complete implementation
of the BLAS interface adapted to asymmetric ARM big.LITTLE processors, 
analyzing the trade-offs between
performance and energy efficiency when compared to existing homogeneous (symmetric) 
multi-threaded BLAS implementations.
Our experimental results reveal important gains in performance while maintaining the energy
efficiency of homogeneous solutions by efficiently exploiting all the resources of the asymmetric
processor.

\end{abstract}



\section{Introduction}

The decay of Dennard scaling~\cite{Den74} during the past decade
marked the end of the ``GHz race'' and the shift towards multicore designs 
due to their more favorable performance-energy ratio.
In addition, the doubling of transistors on chip with each new semiconductor generation,
dictated by Moore's law~\cite{Moo65}, has only exacerbated the 
{\em power wall} problem~\cite{Dur13,Lec13,Luc14}, leading to
the arise of ``dark silicon''~\cite{Esm11} 
and the deployment of heterogeneous facilities for high performance computing.

Asymmetric multicore processors (AMPs) are a particular class 
of heterogeneous architectures equipped 
with cores that share the same instruction set architecture\footnote{According
to this definition, servers 
equipped with one (or more) general-purpose multicore processor(s) 
and a PCIe-attached graphics accelerator, or systems-on-chip like the NVIDIA
Tegra TK1, are excluded from this category.} but
differ in performance, complexity, and power consumption. 
AMPs have recently received considerable attention 
as a means to improve the performance-energy ratio of
computing systems~\cite{Kum04,Hil08,Mor06,Win10}, 
mainly by exploiting the presence of serial and parallel phases within
applications. 

\begin{figure*}[t]
\centering
\begin{minipage}[c]{0.9\textwidth}
\footnotesize
\begin{tabular}{llll}
Loop 1 &{\bf for} $j_c$ = $0,\ldots,n-1$ {\bf in steps of} $n_c$\\
Loop 2 & \hspace{3ex}  {\bf for} $p_c$ = $0,\ldots,k-1$ {\bf in steps of} $k_c$\\
&\hspace{6ex}           \textcolor{darkblue}{$B(p_c:p_c+k_c-1,j_c:j_c+n_c-1)$} $\rightarrow \textcolor{darkblue}{B_c}$ & & // Pack into $B_c$\\
Loop 3 & \hspace{6ex}           {\bf for} $i_c$ = $0,\ldots,m-1$ {\bf in steps of} $m_c$\\
&\hspace{9ex}                     \textcolor{darkred}{$A(i_c:i_c+m_c-1,p_c:p_c+k_c-1)$} $\rightarrow \textcolor{darkred}{A_c}$ & & // Pack into $A_c$ \\
\cline{2-4} 
Loop 4&\hspace{9ex} {\bf for} $j_r$ = $0,\ldots,n_c-1$ {\bf in steps of} $n_r$  & & // Macro-kernel\\
Loop 5&\hspace{12ex}   {\bf for} $i_r$ = $0,\ldots,m_c-1$ {\bf in steps of} $m_r$\\
\cline{2-3}
&\hspace{15ex}             \textcolor{darkgreen}{$C_c(i_r:i_r+m_r-1,j_r:j_r+n_r-1)$} & // Micro-kernel \\
&\hspace{19ex} ~$\mathrel{+}=$     ~\textcolor{darkred}{$A_c(i_r:i_r+m_r-1,0:k_c-1)$} \\
&\hspace{19ex} ~~~$\cdot$\!~~~~\textcolor{darkblue}{$B_c(0:k_c-1,j_r:j_r+n_r-1)$} \\
\cline{2-3}
&\hspace{12ex} {\bf endfor}\\
&\hspace{9ex} {\bf endfor}\\
\cline{2-4} 
&\hspace{6ex} {\bf endfor}\\
&\hspace{3ex} {\bf endfor}\\ 
&{\bf endfor}\\ 
\end{tabular}
\end{minipage}
\caption{High performance implementation of \gemm in BLIS. In the code, $C_c \equiv C(i_c:i_c+m_c-1,j_c:j_c+n_c-1)$
is just a notation artifact, introduced to ease the presentation of the algorithm, while $A_c,B_c$ correspond to actual buffers that are involved in data copies.}
\label{fig:gotoblas_gemm}
\end{figure*}

In this paper we investigate the practical performance-power-energy balance of 
ARM's asymmetric big.LITTLE technology, employing as a case of study the compute-intensive general matrix multiplication (\gemm):
$C \mathrel{+}= A \cdot B$, where the sizes of $A$, $B$, $C$ are
respectively $m \times k$, $k \times n$, $m \times n$.
Most previous related work targets the parallelization of \gemm on 
{\em i)} distributed-memory heterogeneous architectures (see~\cite{Cla11,Bea14} and
references therein); or
{\em ii)} asymmetric multicores, 
but using trivial (unoptimized) implementations of \gemm~\cite{Lak08,Lak09}.
Compared with these other efforts, our paper makes the following contributions:
First, we leverage a static mapping of threads and 
we propose a 
workload partitioning strategy of the BLIS implementation of \gemm specifically tailored for the 
Exynos 5422 big.LITTLE architecture, a system-on-chip (SoC) featuring two processing clusters: an ARM Cortex-A15 
quad core and a Cortex-A7 quad core.  
Second, we perform a detailed evaluation of our solution in terms 
of performance compared with that of the symmetric counterpart on each of the processing clusters of the Exynos 5422.
Third, we perform an energy efficiency evaluation of each approach
 using the GFLOPS/W metric (equivalent to billions of
floating-point arithmetic operations, or flops, per Joule).


\section{Matrix Multiplication for\\ General-Purpose Processors}
\label{sec:blis}

Modern implementations of \gemm for general-purpose architectures,
including BLIS and OpenBLAS,
follow the approach pioneered by GotoBLAS~\cite{Goto:2008:AHP}.
Concretely, BLIS implements \gemm 
as three nested loops around a {\em macro-kernel} plus two packing routines
(see Loops~1--3 in Figure~\ref{fig:gotoblas_gemm}).
The macro-kernel is then implemented in terms of two additional loops around a {\em
micro-kernel} (Loops~4 and~5 in Figure~\ref{fig:gotoblas_gemm}). 
In BLIS, the micro-kernel is typically implemented as a loop around a 
rank--1 (i.e., outer product) update using assembly or
with vector intrinsics, while the remaining five loops are implemented in C;
see~\cite{BLIS1} for further details.
Furthermore,
the BLIS (cache) optimization parameters
$n_c$, $k_c$, $m_c$, $n_r$ and $m_r$ 
are adjusted taking into account the latencies of the floating-point units (FPUs),
number of vector registers, and size/associativity degree of the cache levels. The
goal is that $A_c$ and a narrow column panel of $B_c$, say $B_r$, are feed into
the floating-point units from the L2 and L1 caches, respectively, and these
transfers are fully amortized with enough computation from within
the micro-kernel; see~\cite{BLIS4}.

The parallelization of \gemm in BLIS
is analyzed in~\cite{BLIS2} for conventional multi-threaded
processors and~\cite{BLIS3} for extreme many-threaded architectures
such as the IBM PowerPC A2 (16 cores/64 threads) and the Intel Xeon Phi (60 cores/240 threads).
Basically, in both ``types'' of architectures, the parallel implementations exploit
the concurrency available in
the nested 5--loop organization of the matrix multiplication algorithm at one or multiple levels
(i.e., loops). In general,
the approach takes into account the cache organization of the processor (e.g.,
the presence of multiple sockets, which cache levels are shared/private, etc.), 
while discarding the parallelization of loops that would incur into race conditions
in the update of $C$ as well as loops with too fine granularity.
These analyses~\cite{BLIS2,BLIS3} can be summarized as follows:
\begin{itemize}
\item Parallelization of Loop~5 (indexed by $i_r$). With this option,
      different threads execute different instances of the micro-kernel.
      Furthermore, they access the same column block $B_r$ (of 
      $n_r$ columns) in the L1 cache. 
      The amount of parallelism in this case,
      $\lceil \frac{m_c}{m_r} \rceil$, 
      is limited as $m_c$ is usually a few hundreds. 
\item Parallelization of Loop~4 (indexed by $j_r$).
      Different threads
      access the same block $A_c$, of dimension $m_c \times k_c$, in the L2 cache.
      The time spent in this loop amortizes the cost of packing (moving)
      the block of $A_c$ from main memory into the L2 cache.
      The amount of parallelism, $\lceil \frac{n_c}{n_r} \rceil$,
      is in general larger than in the previous case, 
      as $n_c$ is frequently in the order of several hundreds up to 
      a few thousands.
\item Parallelization of Loop~3 (indexed by $i_c$).
      Each thread packs a different block $A_c$ into the L2 cache
      and executes a different instance of the macro-kernel.
      The number of iterations of this loop is not
      limited by the blocking sizes, but instead depends on the problem dimension $m$. 
      When $m$ is less than the product of $m_c$ and the degree of parallelization of the loop, 
      the blocks $A_c$ will be smaller than the optimal dimension and performance may suffer.
      When there is a shared L2 cache,
      the size of the blocks $A_c$ will have to be reduced by a 
      factor equal to the degree of parallelization of this loop. 
      However, reducing $m_c$ is equivalent to parallelizing the
      first loop around the micro-kernel.
\item Parallelization of Loop~2 (indexed by $p_c$). This is not a good option
      because multiple threads simultaneously update the same parts of 
      $C$, requiring a mechanism to deal with race conditions. 
\item Parallelization of Loop~1 (indexed by $j_c$).
      From a data-sharing perspective, this option is
      equivalent to gaining parallelism outside of BLIS.
      In any case, this parallelization is reasonable on a 
      multi-socket system where each CPU has a separate LLC (last-level cache). 
\end{itemize}

To sum up, these are general guidelines to decide which loops are theoretically good candidates to be parallelized in order
to fully exploit the cache hierarchy of a target architecture. At a glance, the combination of loops to parallelize strongly depends on which
cache(s) are shared. Usually, Loop~1 ($j_c$) is a good candidate when 
the LLC is separated for each CPU (e.g., a multi-socket platform with on-chip L3 cache); 
Loop~3 ($i_c$) should be parallelized when 
each core has its own L2 cache; and
Loops~4 and/or~5 ($j_r$ and $i_r$, respectively) 
are to be parallelized when the cores share the L2 cache.

\section{Matrix Multiplication on AMPs}
\label{sec:performance}

\begin{figure}[th!]
\begin{center}
\includegraphics[width=\columnwidth,height=4.2cm]{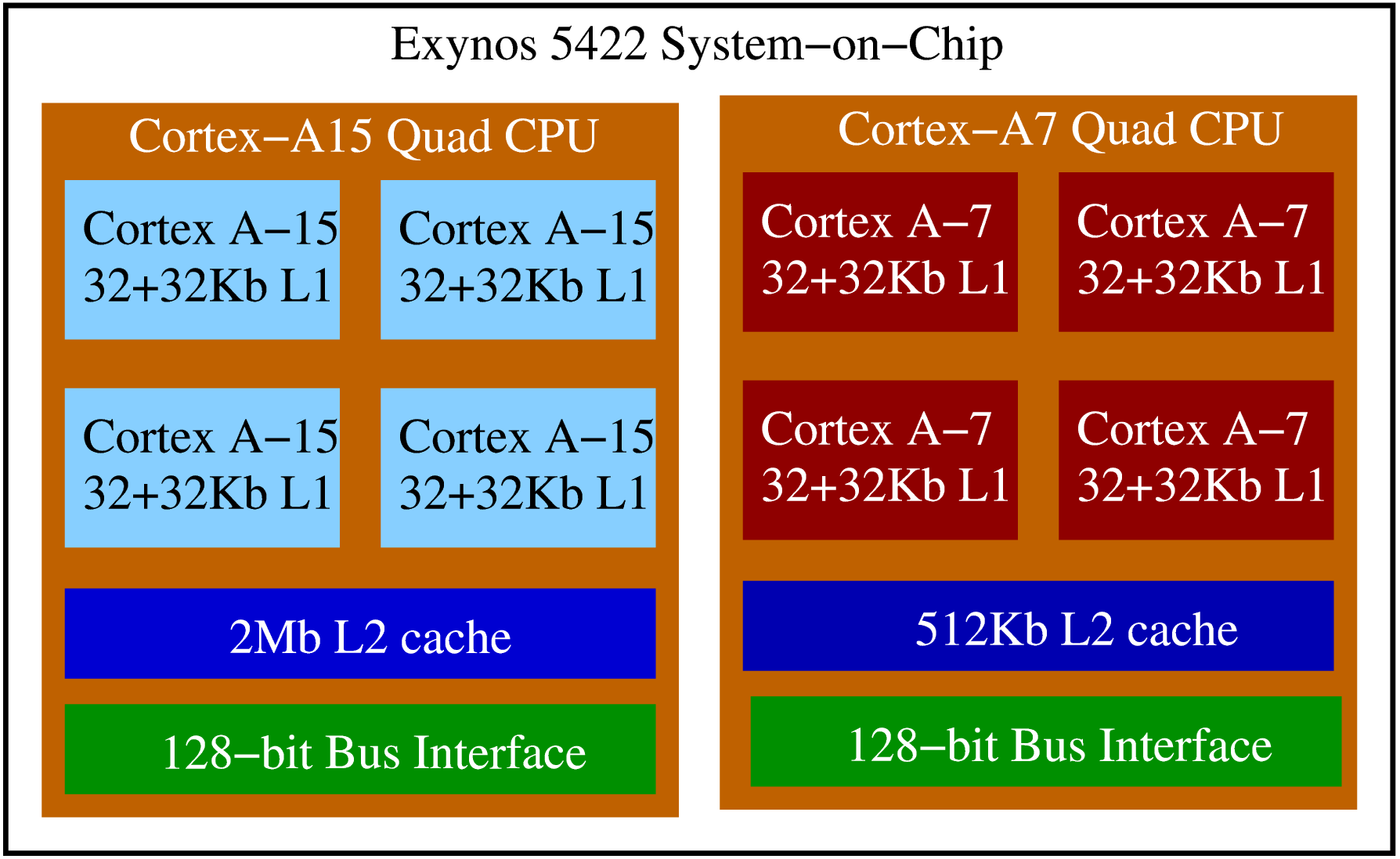}
\end{center}
\caption{\label{fig:exynos} Exynos 5422 block diagram.}
\end{figure}

The ODROID-XU3 contains a Samsung Exynos 5422 SoC with an ARM
Cortex-A15 quad-core processing cluster (running at 1.6~GHz in our setup) and a Cortex-A7 quad-core processing 
cluster (at 1.3~GHz).
Both clusters access a shared DDR3 RAM (2~Gbytes) via 128-bit coherent bus interfaces.
Each ARM core (either Cortex-A15 or Cortex-A7) has a 32+32-Kbyte L1 (instruction+data) cache.
The four ARM Cortex-A15 cores share a 2-Mbyte L2 cache, while the four ARM Cortex-A7 cores
share a smaller 512-Kbyte L2 cache; see Figure~\ref{fig:exynos}.

%

In order to attain high performance, a preliminary step is
to determine the optimal block sizes ($m_c$, $k_c$, $n_c$) 
for the target architecture and precision 
(all our experiments use {\sc ieee} 754 double-precision arithmetic). 
For this purpose, we performed an empirical search 
on the Cortex-A15 cores, detecting the optimal values at
$m_c = 176$ and $k_c = 368$. In this architecture, $n_c$ plays a minor role and is simply set to
$n_c = 4,096$ ($n_c$ is usually related to L3 cache, which is not present on these ARM CPUs). 
The micro-kernel for this architecture is hand-coded with $m_r = 4$ and $n_r = 4$. 
These optimal values are used in this work for both the Cortex-A7 and the Cortex-A15 cores.

\subsection{Mapping multi-threaded BLIS to AMPs}

BLIS allows to select, at run time, which 
(one or more) of the five internal loops are parallelized. 
In particular, if one of the loops is parallelized, a static partition and 
mapping of loop iteration chunks to the 
OpenMP threads is performed prior to the beginning of the loop.

Our asymmetric version of BLIS integrates the following three new features,
which modify the behavior of the multi-threaded BLIS at run time, in
order to 
accomodate an AMP architecture: 
{\em i)} a mechanism to create ``slow'' and ''fast'' threads, which will be bound
upon initialization of the library to LITTLE (Cortex-A7) and big (Cortex-A15) cores; 
{\em ii)} a mechanism to decide which one of the loops that are parallelized
needs to be partitioned and assigned to slow/fast cores asymmetrically 
(thus, chunks assigned to threads will no longer be of uniform size, 
but partitioned according to the capabilities of each type of core); and 
{\em iii)} an interface to specify the ratio of performance between LITTLE and big cores, 
which will ultimately define the number of iterations assigned to each thread/core. 
All these mechanisms are currently modified via environment variables, 
but the development of an {\em ad-hoc} API is part of ongoing work. 

For the target Exynos 5422 SoC, given the memory organization of the 
this big.LITTLE architecture (private L1 cache per core, shared L2 cache 
per cluster, lack of L3 cache), and the guidelines given for the parallelization of BLIS
\gemm at the end of section~\ref{sec:blis}, we chose the approach explained next for the
parallelization on the target Exynos 5422 AMP.  

\begin{figure}[th!]
\begin{center}
\includegraphics[width=\columnwidth,height=5.5cm]{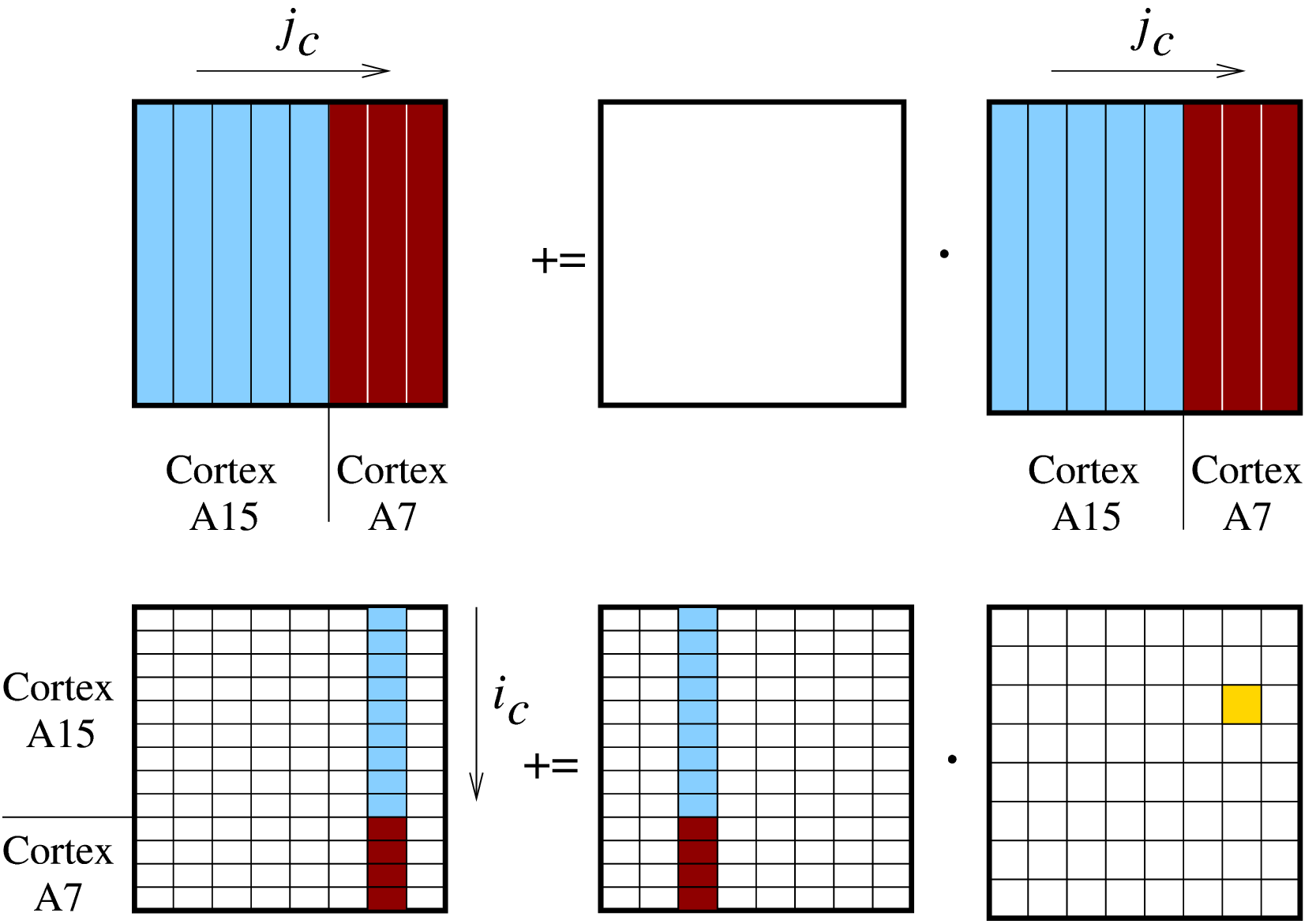}
\end{center}
\caption{\label{fig:A15vsA7} Workload distributions 
         for the matrix multiplication $C \mathrel{+}= A \cdot B$
         between the A15 and A7 quad-core clusters. 
         Top: parallelization of Loop~1 ($j_c$); 
         bottom: parallelization of Loop~3 ($i_c$).
         In the bottom plot, the small rectangles, delimited by the fine lines,
         denote the operands of the macro-kernel 
         $\textcolor{darkgreen}{C_c} \mathrel{+}= \textcolor{darkred}{A_c} \cdot 
         \textcolor{darkblue}{B_c}$.}
\end{figure}

      At a coarse-grain, 
      the computational workload of the complete multiplication $C \mathrel{+}= A \cdot B$
      is distributed among the Cortex-A15 and Cortex-A7 clusters by parallelizing
      either Loop~1 ($j_c$) or~3 ($i_c$). 
      In order to preserve the optimal cache parameters during the execution of \gemm,
      while attaining a distribution of the workload proportional to computational
      power of the A15 vs A7 clusters, we assign
      a different number of iterations of the parallelized loop to each cluster;
      see, e.g., Figure~\ref{fig:A15vsA7}.
      In particular, the ratio applied to distribute the iteration space
      between the Cortex-A15 and Cortex-A7 for \gemm\ has been empirically determined to be 6:1\footnote{This ratio varies depending on the target architecture, 
     core operating frequency, and specific routine, so it should be adjusted accordingly.}.

\begin{figure}[th!]
\begin{center}
\includegraphics[width=\columnwidth,height=5.5cm]{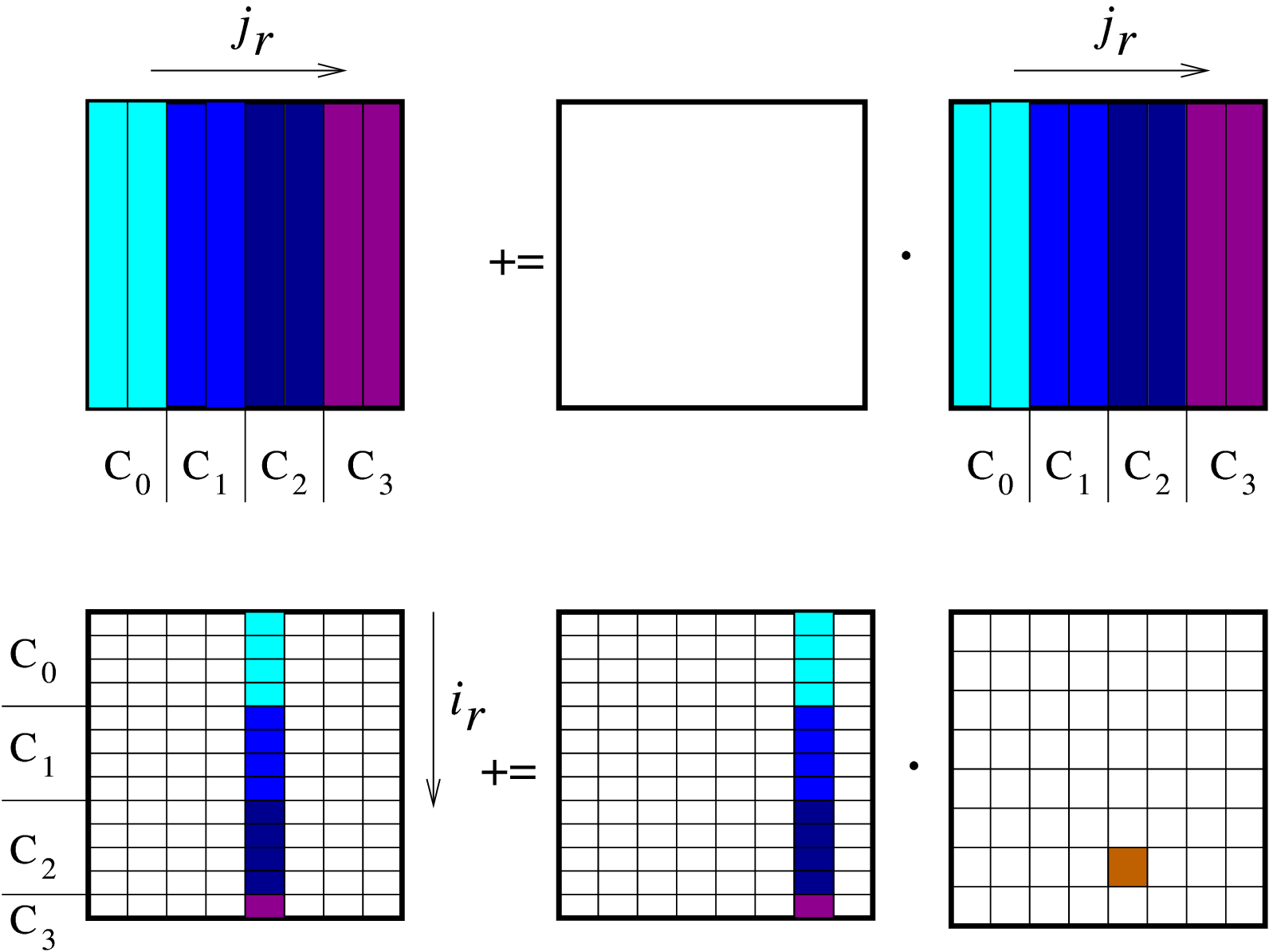}
\end{center}
\caption{\label{fig:Cores4} Workload distributions 
         for the macro-kernel multiplication 
         $\textcolor{darkgreen}{C_c} \mathrel{+}= \textcolor{darkred}{A_c} \cdot 
         \textcolor{darkblue}{B_c}$
         between four cores of the same type 
         (C$_{\mbox{\rm 0}}$, C$_{\mbox{\rm 1}}$, C$_{\mbox{\rm 2}}$, C$_{\mbox{\rm 3}}$). 
         Top: parallelization of Loop~4 ($j_r$); 
         bottom: parallelization of Loop~5 ($i_r$). In this example,
         the OpenMP chunk size equals~2 in the first case and~4 in the second.}
\end{figure}

      At a finer-grain, the execution of each macro-kernel 
      $\textcolor{darkgreen}{C_c} \mathrel{+}= \textcolor{darkred}{A_c} \cdot \textcolor{darkblue}{B_c}$ (see Figure~\ref{fig:gotoblas_gemm})
      is partitioned among the cores of the same 
      type by parallelizing Loops~4 ($j_r$), 5 ($i_r$) or both; see, 
      e.g., Figure~\ref{fig:Cores4}.

\section{Evaluation of Performance and Energy Efficiency}
\label{sec:results}

The goal of the performance and energy efficiency tests in this section is to carry out 
an experimental study of both metrics comparing
the original multi-threaded of \gemm in BLIS against our asymmetric-aware implementation.
%
In all tests, we ensure the cores 
run at their highest frequency by setting the {\em performance} governor. 
Codes are instrumented with the {\tt pmlib}~\cite{AlonsoICPP12} framework, which 
collects power consumption data corresponding 
to instantaneous power readings from four independent sensors in the board 
(for the Cortex-A7 cores, Cortex-A15 cores, DRAM and GPU),
with a sampling rate of 200 ms.

\begin{figure}[!ht]
\includegraphics[width=\columnwidth,height=5cm]{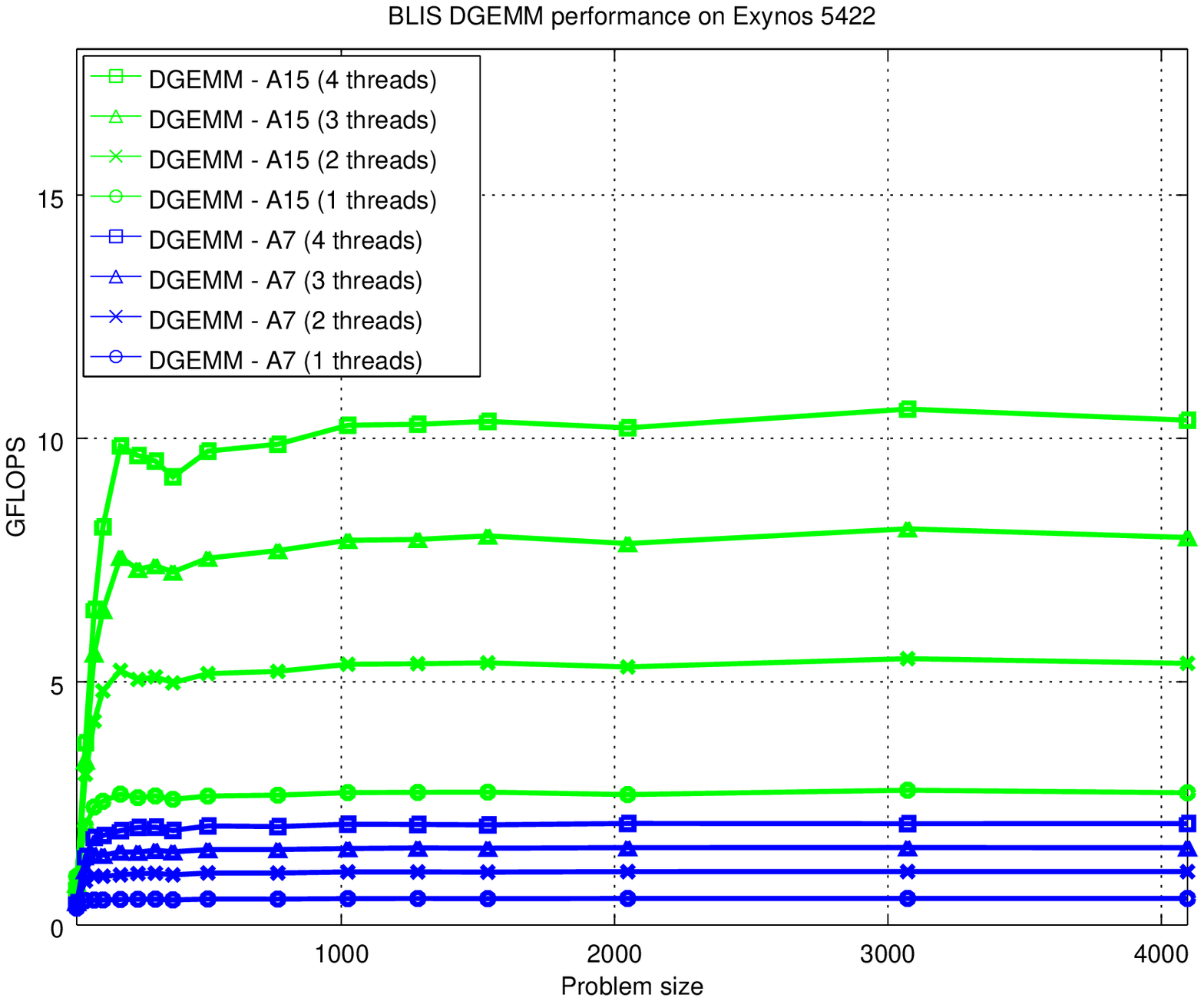}
\includegraphics[width=\columnwidth,height=5cm]{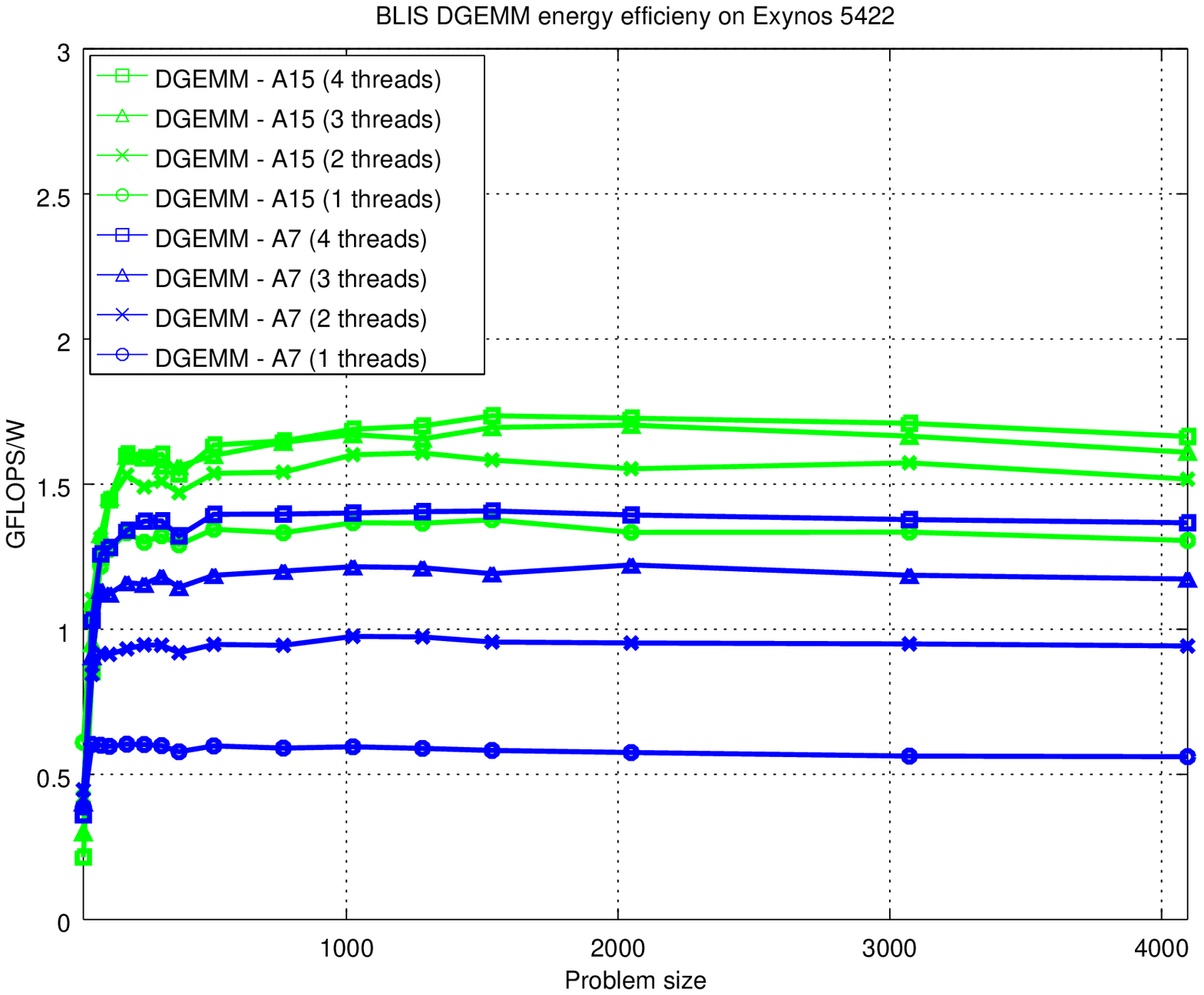}
\caption{\label{fig:GFLOPS}Performance (top) and energy efficiency (bottom) of the BLIS DGEMM using 
         exclusively one type of core, for a varying number of threads.}
\end{figure}

The first round of experiments analyzes the performance and energy behavior 
of the Cortex-A7 and the Cortex-A15 core types when working in isolation. For this purpose,
we execute a collection of \gemm kernels using one of the fine-grain parallelization 
exposed in Section~\ref{sec:performance}. Concretely, as the L2 cache
is shared among the cores of a cluster, we parallelize Loop~4 
using 1, 2, 3 and 4 threads (cores), with the performance and energy results
in Figure~\ref{fig:GFLOPS}. These plots reveal that the Cortex-A15 cores 
clearly deliver higher performance, with a rough increase of 2.5 GFLOPS per core, 
attaining a peak performance of about $10.2$ GFLOPS with 4 threads. 
For the Cortex-A7 cores, the 
performance peaks are around $2.0$ GFLOPS and is also attained with 4~cores. 
Regarding energy efficiency, the Cortex-A15 obtains 
the best results in terms of GFLOPS/W. However, the benefits from
increasing the number of threads in this case are less significant (0.055 GFLOPS/W per core) 
when compared with those obtained with the Cortex-A7 cores (0.193 GFLOPS/W per core). 
It is also worth emphasizing that the use of 4 Cortex-A7 cores is more energy-efficient 
than an alternative that leverages a single Cortex-A15 core, though the overall 
performance of the former is slightly worse.

The second round of experiments 
evaluates the performance and energy efficiency of the asymmetric-aware port of BLIS to the 
big.LITTLE architecture. For this purpose,
we run a collection of \gemm kernels, 
relaying on a 2-way parallelization to distribute iterations of Loop~3
(see Section \ref{sec:performance}),
with a ratio of 6:1, among the cores of the fast and slow clusters, and taking advantage of
the independent L2 cache per cluster in this manner.  
For the fine-grain parallelization, 4 threads are leveraged 
in order to assign chunks of the iteration space 
for Loop~4 to each core within the cluster. Our experiments with different
configurations revealed this option to be the most efficient 
for the target big.LITTLE architecture.

\begin{figure}[!ht]
\includegraphics[width=\columnwidth,height=5cm]{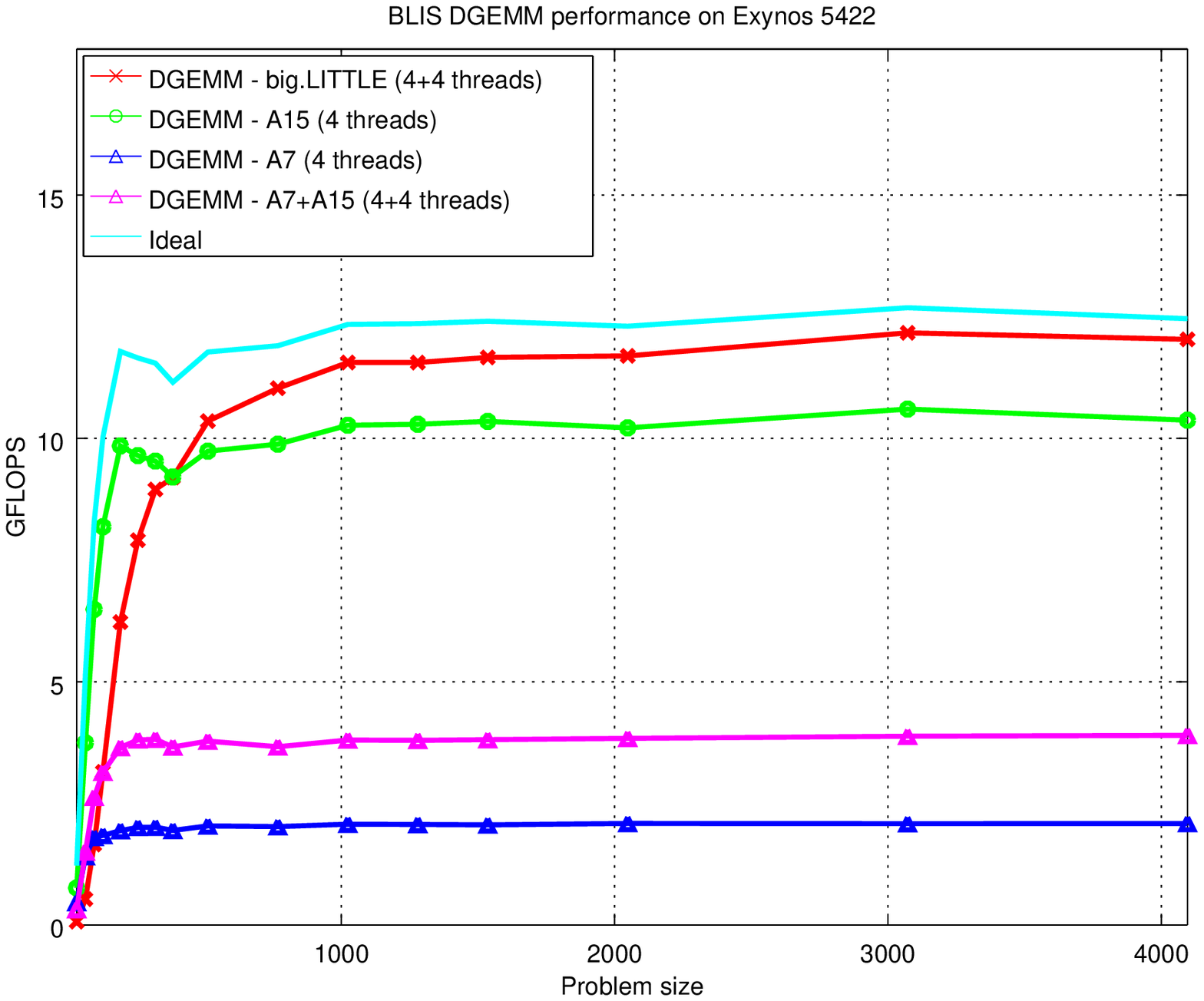}
\includegraphics[width=\columnwidth,height=5cm]{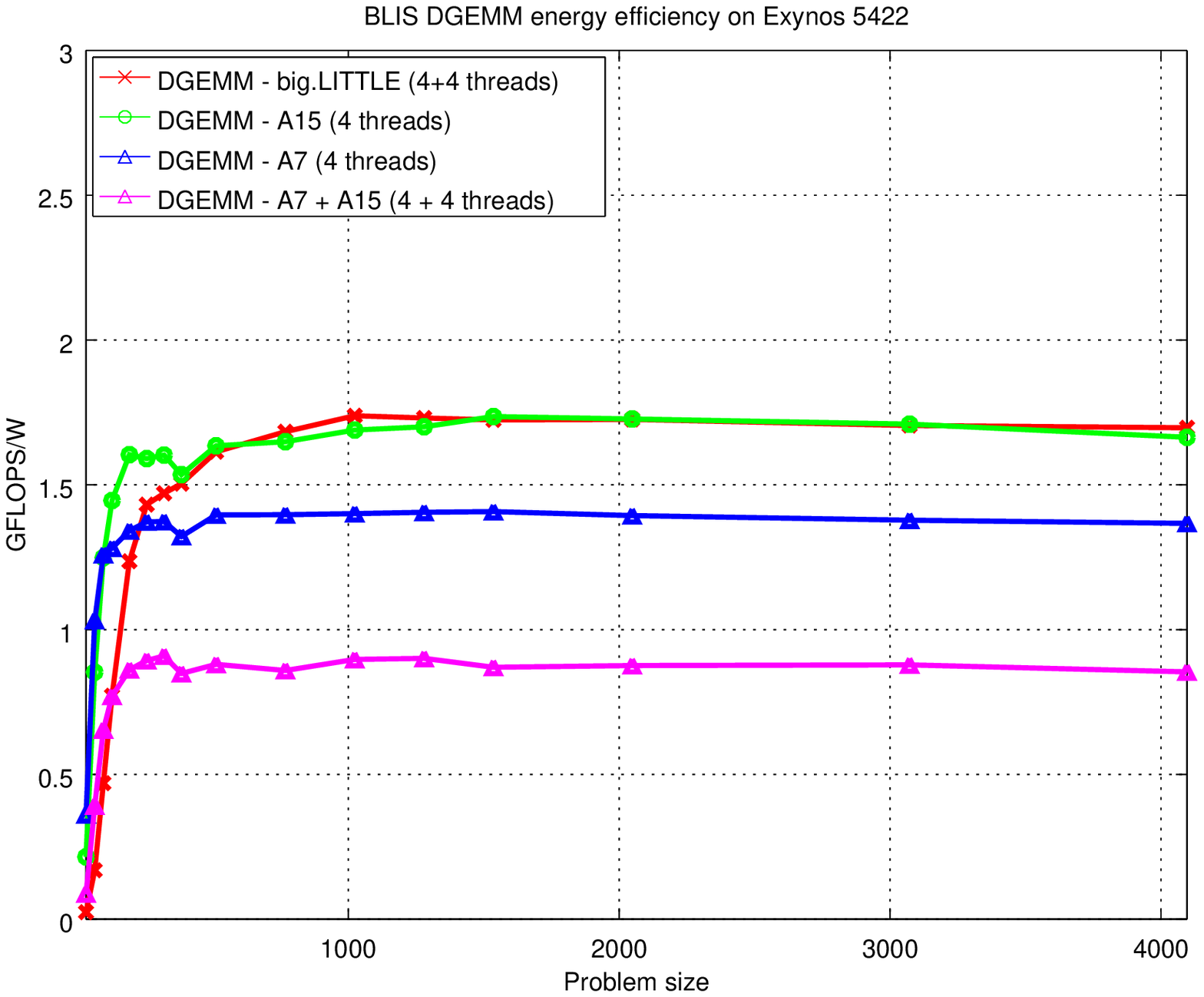}
\caption{\label{fig:AMP}Performance (top) and energy efficiency (bottom) of the BLIS DGEMM 
         implementations using a single as well as different types of cores.}
\end{figure}

Figure~\ref{fig:AMP} reports the results for this second evaluation.
The line labeled as ``{big.LITTLE (4+4 threads)}'' 
corresponds to the asymmetric-aware implementation. 
The same \gemm kernels were computed with BLIS using a symmetric workload distribution 
(the iteration space is equally distributed among the Cortex-A7 and Cortex-A15 cores), 
with the results labelled as ``{A7+A15 (4+4 threads)}'' in the figure. 
For comparison purposes, the performance and energy obtained using exclusively 
four Cortex-A7 or four Cortex-A15 CPUs are also added. Finally, the ``ideal'' line
corresponds to the sum of the peak performances of the configurations 
that use four cores of each of the two types in isolation (i.e., the performance of the four
Cortex-A15 cores plus the performance of the four Cortex-A7 cores).

These performance results show that the AMP configuration 
outperforms the peak performance of all other configurations being close to the ideal case. 
The increment 
compared to the configuration that employs four Cortex-A15 cores for the largest tested problem
is close to 20\%. 
The asymmetric version does not outperform the original version for small matrices 
though, as the chunks assigned to the big and LITTLE cores are, in those cases, 
too small to exploit the asymmetric architecture.
In terms of energy-efficiency, the AMP configuration is as efficient as the symmetric setup using 
exclusively four Cortex-A15 CPU.

The symmetric workload distribution attains about 40\%
of the highest performance that is observed when employing only the Cortex-A15 cores. 
The reason is that, with the symmetric workload distribution, 
thread scheduling is delegated to the operating system or the
OpenMP runtime, using a homogeneous distribution of chunks. 
This causes a severe load imbalance as the fast Cortex-A15 
threads finish processing their assigned chunk, and have to wait a long time
for the Cortex-A7 threads to complete their assignment. 
The energy-efficiency is also affected,
and this configuration achieves the worst energy-efficiency.

Diving into details that explain the energy efficiency of our implementations, 
Table~\ref{tab:power_detail} shows a breakdown of power/energy per component
of the SoC, for a particular problem size: $m=n=k=4,096$. 
This table shows the (average) power consumption and energy efficiency when 
employing {\em i)} from~1 to~4 threads of a single cluster; 
{\em ii)} the AMP configuration with all 4+4 cores; and 
{\em iii)} the symmetric configuration of BLIS using all 4+4 cores. 
The first four columns report the average power consumption gathered from the SoC sensors, 
while the average power consumption of the entire SoC is in the fifth column. 
The performance achieved by the different configurations is reported in the 
sixth column and the energy efficiency is displayed in the last one.

The first aspect to note is that, as expected, the Cortex-A15 cores dissipate more power 
than the Cortex-A7 cores. Indeed, a single Cortex-A15 core roughly doubles the power 
dissipation rate of four combined Cortex-A7 cores, and the Cortex-A15 CPU in idle state consumes 
more power than two Cortex-A7 cores in execution. A second issue is that the 
memory (DRAM) and total power consumption of the AMP and symmetric configurations are close to
those obtained by adding the corresponding values of the two CPU clusters in isolation. 
An exception is the total power consumption with the symmetric configuration, in which a
significant decrease is observed due to the Cortex-A15 cores completing their share
of the work much earlier than the Cortex-A7 cores. 
This aspect strongly affects the energy efficiency of the symmetric configuration 
as the power consumption is three times higher than that obtained with the entire Cortex-A7 cluster,
but the performance is only doubled. As expected, the AMP configuration is the one 
that dissipates a higher power rate, as it fully utilizes all the available resources. 
On the other hand, it also obtains the shortest execution time, 
yielding the best energy-to-solution.

\begin{table*}
\begin{center}
\begin{small}
\begin{tabular}{|c||c|c|c|c|c||c|c|}
  \hline
  \multirow{2}{*}{Configuration}   & \multicolumn{5}{c||}{Average Dissipated Power (W)} & \multirow{2}{*}{GFLOPS}   & \multirow{2}{*}{GFLOPS/W} \\ 
                  & A7       & A15       & DRAM        & GPU       & Total       &          &          \\ \hline \hline
  Asymmetric BLIS & 0.785    & 5.994     & 0.191      & 0.119     & 7.091       & 12.035   & 1.697    \\ \hline \hline
  1xA15           & 0.109    & 1.828     & 0.060      & 0.083     & 2.081       & ~2.718   & 1.305    \\ \hline
  2xA15           & 0.124    & 3.242     & 0.076      & 0.099     & 3.543       & ~5.377   & 1.517    \\ \hline
  3xA15           & 0.135    & 4.613     & 0.091      & 0.106     & 4.946       & ~7.963   & 1.609    \\ \hline
  4xA15           & 0.140    & 5.878     & 0.105      & 0.110     & 6.233       & 10.374   & 1.664    \\ \hline \hline
  1xA7            & 0.305    & 0.499     & 0.066      & 0.102     & 0.973       & ~0.546   & 0.560    \\ \hline
  2xA7            & 0.488    & 0.501     & 0.072      & 0.102     & 1.164       & ~1.098   & 0.942    \\ \hline
  3xA7            & 0.661    & 0.503     & 0.084      & 0.103     & 1.352       & ~1.587   & 1.173    \\ \hline
  4xA7            & 0.831    & 0.502     & 0.089      & 0.103     & 1.526       & ~2.086   & 1.366    \\ \hline \hline
  Symmetric BLIS  & 0.810    & 3.440     & 0.201      & 0.109     & 4.562       & ~3.897   & 0.854    \\ \hline
\end{tabular}
\caption{\label{tab:power_detail}Power consumption breakdown and energy efficiency for DGEMM ($m=n=k=4096$) on the Exynos 5422 SoC, using
different thread configurations. The rows labeled as {\it Asymmetric BLIS} and {\it Symmetric BLIS} use all the available eight cores in the
SoC, using our modified BLIS version and the original BLIS multi-threaded implementation, respectively.}
\end{small}
\end{center}
\end{table*}

\section{Conclusions}
\label{sec:conclusions}

In this paper, we have proposed several mechanisms to map the high-performance
multi-threaded implementation of the matrix multiplication in the BLIS library 
to an asymmetric ARM big.LITTLE 
(Cortex A15+A7) SoC. Our results reveal excellent improvements in performance 
compared with a homogeneous implementation that operates exclusively on one type of core 
(either A15 or A7), and also with respect to multi-threaded implementations that
rely on a symmetric work distribution and delegate scheduling to the operating system.

This is the first step towards a full BLAS implementation optimized for 
big.LITTLE architectures, which is the ultimate goal of our work. 
We believe that the approach applied to \gemm carries over
to the rest of the BLAS. However, there are still a number of issues 
that need to be addressed to further increase performance and adaptation to the architecture. 
Among those, the most significant ones are the integration of different micro-kernels and 
block sizes tuned to each type of core in order to extract the maximum performance,
and the dynamic distribution and mapping of the 
workload to each type of core transparently to the programmer. A port to a 64-bit 
ARMv8 architecture, and performing a experimental study on
architectures with different number of big/LITTLE cores are also key milestones in our roadmap.

\section*{Acknowledgments}

The researchers from Universitat Jaume~I 
were supported by project CICYT TIN2011-23283 of
MINECO and FEDER, the EU project FP7 318793 ``EXA2GREEN'' and the FPU
program of MECD.
The researchers from Universidad Complutense de Madrid 
were supported by project CICYT TIN2012-32180.

\bibliographystyle{plain}
\bibliography{enrique,energy,asymmetric}

\end{document}